\documentclass[preprint, pra]{revtex4}
\usepackage{amsmath}
\usepackage{amssymb}
\usepackage{graphicx}
\usepackage{enumitem}

\usepackage{bm}
\renewcommand{\vec}[1]{\boldsymbol{\mathbf{#1}}}

\newcommand{\subfigimg}[3][,]{%
  \setbox1=\hbox{\includegraphics[#1]{#3}}
  \leavevmode\rlap{\usebox1}
  \rlap{\hspace*{-10pt}\raisebox{\dimexpr\ht1-2\baselineskip}{#2}}
  \phantom{\usebox1}
}
\usepackage[font={footnotesize}]{caption}

\begin{document}

\title{Solitons in a continuous classical Haldane-Shastry spin chain}
 
\author{Tianci Zhou}
\email{tzhou13@illinois.edu}
\affiliation{University of Illinois, Department of Physics, 1110 W. Green St. Urbana, IL 61801 USA}
 
\author{Michael Stone}
\email{m-stone5@illinois.edu}
\affiliation{University of Illinois, Department of Physics, 1110 W. Green St. Urbana, IL 61801 USA}

\date{\today}

\begin{abstract}
Motivated by Polychronakos' discovery that solitons exist in the hydrodynamic equations of continuum version of the Calogero model, we seek solitons in the classical dynamics of a continuum version of the Haldane-Shastry spin chain. We have obtained analytic multi-lump solitary wave solutions for our spin-field equation, and these solutions possess interesting topological features. We have performed numerical collision experiments showing that these solitary waves survive collisions, and thus suggest the existence of true multi-soliton solutions.
\end{abstract}
  
\maketitle

\section{Introduction and Motivation}

Solitons are localized wave packets that survive unchanged through collisions.  Since their accidental discovery as water waves described by the KdV equation (See \cite{filippov_great_2010} for an account of this), soliton solutions have been found in many systems of partial differential equations. They have also been observed in many physical systems --- for example, nonlinear optics\cite{stegeman_optical_1999,bjorkholm_cw_1974}, matter wave solitons in Bose-Einstein condensation\cite{strecker_formation_2002}, and vortex rings in ferromagnetic materials\cite{cooper_propagating_1999, sutcliffe_vortex_2007, niemi_leapfrogging_2014}. 

The soliton property is closely related to the integrability of the underlying system of differential equations. Classical integrability ensures that there are as many Poisson-commuting integrals of motion as degrees of freedom, and the associated conservation laws suppress the available post-collision phase space to the extent that the only possible outgoing solutions are non-diffractive and just a permutation of the incoming solitary waves. 

It is natural to describe a quantum integrable system as one with as many mutually commuting local operators as degrees of freedom. This is not necessarily the most useful definition as knowing such operators does not always help solve the system. Sutherland therefore defines a {\it quantum integrable\/}  system as one that supports non-diffractive scattering\cite{sutherland_beautiful_2004}. A beautiful example is provided by the Calogero-Sutherland models which are both classical and quantum integrable \cite{polychronakos_physics_2006}. It is possible to take a continuum limit of the classical Calogero model and the resulting hydrodynamic equations of motion possess multi-soliton solutions\cite{polychronakos_waves_1995, andric_solitons_1995}. This remarkable result suggests that the continuum model remains classically integrable. 

With Calogero hydrodynamics as motivation, we here construct a classical version of Haldane-Shastry (HS) spin chain\cite{haldane_exact_1988,shastry_exact_1988, haldane_spinon_1991}.  The HS spin chain is the infinite mass limit of the quantum integrable {\it spin\/}-Calogero model\cite{polychronakos_lattice_1993,talstra_integrals_1995,bernard_yang-baxter_1993} and is also quantum integrable\cite{polychronakos_exchange_1992}.

To obtain a classical continuum version of the spin chain, we first interpret the spin-$\frac{1}{2}$ exchange term as ferromagnetic interaction between the spins, and then replace spin-$(j=1/2)$ by a large enough value of $j$ so that the dynamics becomes classical. The degrees of freedom of our classical model are therefore unit vectors $\vec{m}_i$ at equally spaced lattice sites. We take the interactions to be ferromagnetic so that we can anticipate a continuum limit in which ${\bf m}_i\to {\bf m}(x)$ with ${\bf m}(x)$ being a smooth function. The resulting equation of motion for $\vec{m}(x)$ is then a non-local generalization of the (known to be integrable) Landau-Lifshitz equation where the second derivative with respect to $x$ is replaced by the derivative of a Hilbert transform in a manner reminiscent of Benjamin-Ono equation\cite{benjamin_internal_1967,ono_algebraic_1975}(itself an integrable equation having known multi-soliton solutions\cite{joseph_multi-soliton-like_1977}). We will see that our generalized Landau-Lifshitz equation possesses both interesting analytic solutions and numerically-obtained multi-soliton solutions. 

The paper is organized as follows. In section \ref{sec:continuum} we construct our continuum model and derive the classical equation of motion. In section \ref{sec:single-speed} we introduce numerical and analytical methods, and use them to solve the single-speed sector. In section \ref{sec:multi-soliton}, we presents numerical collision experiments of multiple solitons. Section \ref{sec:discussion} is the discussion of the solutions, especially their topological features. \ref{sec:summary} concludes our results. Some technical details are presented in the appendices.

\section{The Classical and Continuum Hamiltonian}
\label{sec:continuum}
\subsection{The Continuum Hamiltonian}

The ferromagnetic version of the original Haldane-Shastry model has Hamiltonian
\begin{equation}
H_{\text{HS}} = \sum_{i<j} \frac{1 - \vec{\sigma}_i \cdot \vec{\sigma}_j  }{(x_i - x_j)^2}.
\end{equation}
Here ${\bm \sigma}_i=(\sigma_x,\sigma_y,\sigma_z)_i$ are Pauli matrices that act on the Hilbert space of a spin at position $x_i$ and we have taken $1 - \vec{\sigma}_i \cdot \vec{\sigma}_j $ in the interaction so that the energy would be zero if all spins were parallel. 

The spin-$1/2$ dynamics is very quantum mechanical. There are two routes to modifying the system so as to obtain a classical limit. One is to replace the ${\rm SU}(2)$ spin group with ${\rm SU}(N)$, while preserving the interpretation as a spin exchange interaction. Thus
\begin{equation}
P_{ij}  = \frac 12( 1 + \vec{\sigma}_i \cdot \vec{\sigma}_j ) \to \frac 1{N} + {\bm \lambda}_i \cdot {\bm \lambda}_j,  
\end{equation}
where ${\bm \lambda}=(\lambda_1,\ldots \lambda_{N^2-1})$ are generators of ${\rm SU}(N)$ normalized so that ${\rm tr}(\lambda_a\lambda_b)=\delta_{ab}$. The classical approximation then has the ${\bm \lambda}_i$ taking values in a suitable co-adjoint orbit.

The other, which is the route we will take, is to regard the ${\bm \sigma}_i\cdot {\bm \sigma}_j$ part of the exchange term as a ferromagnetic interaction between spin-($j=1/2$) particles and then take $j$ large enough for the spins to become classical. The resulting interaction term can no longer be interpreted as a spin exchange, as the general spin-$j$ exchange term is a higher-order polynomial in the spin operators.

Our original HS Hamiltonian has therefore been replaced by
\begin{equation}
\label{eqn-discrete-H}
H_{\text{HS}} \rightarrow H_{\rm classical}=   \sum_{i<j}\frac{1 - \vec{m}_i \cdot \vec{m}_j  }{(x_i - x_j)^2}.
\end{equation}
We place the spin chain on a 1d lattice with spacing $a$, so ${a}^{-1}$ is the density $\rho$ and we can write 
\begin{equation}
H_{\rm classical} = \rho^2 \sum_{i= -\infty}^{\infty}  \vec{m}_i  \cdot  \sideset{}{'}\sum_{j = -\infty}^{\infty} \frac{ \vec{m}_i  -   \vec{m}_j}{ (i-j)^2 }
\end{equation}
We now wish to replace the ${\bf m}_i$ by the smooth function ${\bf m}(x)$ and the sums over $i$ and $j$ by integrals.

There are two ways to approximate sums by integrals, one is to use real-space Euler-Maclaurin summation formula, and the second makes use of Fourier interpolation in momentum space. The two methods give the same result (see appendix \ref{app:hydro-limit}): for slowly varying ${\bf m }_i$
\begin{equation}
H_{\rm classical} \sim  H_0  \stackrel{\rm def}{=}  \rho^2 \int_{-\infty}^{\infty} dx \left\{ \pi\, \vec{m} \cdot \partial_x{\bf m}_{\mathcal{H}}- \frac{1}{2} (\partial_x \vec{m})^2  \right\}.
\end{equation}
Here ${\bf m}_{\mathcal H}(x)$ is the Hilbert transform of ${\bf m}(x)$ defined by the principal-part integral 
\begin{equation}
{\bf m}_{\mathcal H}(x)=  \frac{\rm P}{\pi}\int_{-\infty}^{\infty}  \frac{{\bf m}(\xi)}{x-\xi} d\xi.
\end{equation}
On using the identity 
\begin{equation}
  \frac{d}{dx}\left( \frac{\rm P}{\pi}\int_{-\infty}^{\infty}  \frac{{m}_i(\xi)}{x-\xi} d\xi\right)=   \frac{\rm P}{\pi}\int_{-\infty}^{\infty}  \frac{{m}_i(x)- m_i(\xi)}{(x-\xi)^2} d\xi
 \end{equation} 
together with $|{\bf m}|^2=1$, we see that the first term in the integral is the na\"ive continuum limit where we simply replace each sum by an integration. The second term is a correction that we have kept because such corrections play a vital role in preserving the soliton property in the classical hydrodynamics of the Calogero Sutherland models. It is informative to look at the role of this correction in momentum space. We have 
\begin{eqnarray}
\label{eq:fourier-space}
H_0 &=& \rho^2 \int_{-\infty}^{\infty} \frac{dk}{2\pi}\left\{  -i \text{sgn}(k) \cdot ik \cdot \pi  -\frac{1}{2} k^2 \right\}|\vec{m}(k)|^2 \nonumber\\
&=& \rho^2 \int_{-\infty}^{\infty} \frac{dk}{2\pi}  \left\{  \pi |k|- \frac 12 k^2     \right\} |\vec{m}(k)|^2 .
\end{eqnarray}
We see that the correction term leads to our continuum approximation to the manifestly-positive discrete-model energy density being in danger of becoming negative when it has large $k$ component. Of course the continuum limit is supposed to be smooth and so large $k$ are not supposed to appear. In particular, values of $|k|>\pi$ are meaningless since they alias to $k-2\pi$. Nonetheless, 
using this Hamiltonian leads to instabilities in the numerical simulations. The evolution and gradient-descent mentioned in section \ref{sec:single-speed} all develop unwanted anti-ferromagnetic ($k=\pi$) oscillations. We therefore tentatively discard the double derivative term $- \frac{1}{2} (\partial_x \vec{m})^2$ in the Hamiltonian. 

Apart from the numerical instability, there is a legitimate reason for the discard. In momentum space, the Hilbert transformed term scales as $|k|$ and the double derivative term scales like $k^2$. We only care about the physics in large distance, so $k$ is very close to zero. The Hilbert transformed term is thus far larger than the double derivative term and, in the hydrodynamic limit it is legitimate to neglect the later. The Hamiltonian we actually use is therefore
\begin{equation}
H  =  \frac{1}{2} \int_{-\infty}^{\infty} dx \,  \vec{m} \cdot  \partial_x \vec{m}_{\mathcal{H}}.
\end{equation}
The constants $\rho^2$ and $\pi$ have also been dropped as they only affect the time scale.

\subsection{The Equation of Motion}

The classical motion of a single unit-vector spin ${\bf m}=(m_1,m_2,m_3)$ is derived from the Poisson bracket
\begin{equation}
\{m_i,m_j\}= \epsilon_{ijk}m_k.
\end{equation}
We extend this bracket to functionals $F[{\bf m}]$, $G[{\bf m}]$ of a continuous spin-field ${\bf m}(x)$ by setting 
\begin{equation}
\{m_i(x),m_j(x')\}= \epsilon_{ijk}m_k\,\delta(x-x'),
\end{equation}
and hence \cite{tjon_solitons_1977}, 
\begin{equation}
\{F, G\} = \epsilon_{ijk} \int_{-\infty}^{\infty}  \frac{\delta F}{\delta m_i(x)}  \frac{\delta G}{\delta m_j(x)} m_k(x) dx. 
\end{equation}
The time evolution of $\vec{m}$ for a Hamiltonian $H[{\bf m}]$ is then  
\begin{equation}
\label{equ:H-eqn}
\frac{\partial \vec{m}}{\partial t} = \{ \vec{m}, H \} =  \frac{\delta H}{\delta \vec{m}} \times \vec{m}. 
\end{equation}
For example 
\begin{equation}
H_{LL}[{\bf m}]= \frac{1}{2} \int_{-\infty}^{\infty} dx \,\, (\partial_x \vec{m} )^2  
\end {equation}
gives Landau-Lifshitz equation (LLE) \cite{fogedby_solitons_1980,lakshmanan_fascinating_2011}
\begin{equation}
\frac{\partial {\bf m}}{\partial t}= {\bf m}\times \frac{\partial^2 {\bf m}}{\partial x^2},
\end{equation}
whose soliton solutions are connected to helical curve motion\cite{betchov_curvature_1965,hasimoto_soliton_1972, lamb_solitons_1976,lakshmanan_continuum_1977} and have been extensively studied \cite{zakharov_equivalence_1979,tjon_solitons_1977,takhtajan_integration_1977, lakshmanan_dynamics_1976,bikbaev_landau-lifshitz_2014}.  Many of the methods used for the LLE turn out to be applicable to our problem.

For our classical continuum version of the Haldane-Shastry spin chain, the equation of motion becomes
\begin{equation}
\label{equ:EOM}
 \frac{\partial \vec{m}}{\partial t} = \partial_x\vec{m}_{\mathcal{H}} \times \vec{m }
\end{equation}
Because differentiation with respect to $x$ commutes with taking a Hilbert transform, equation \eqref{equ:travel-wave} can also be written as
\begin{equation}
 \frac{\partial  \vec{m}}{\partial t } = \left( \frac{\partial  \vec{m}}{\partial x}\right)_{\mathcal{H}}\times {\bf m}.
\end{equation}

\subsection{Conserved Quantities}

Integrable systems with continuous degrees of freedom process infinitely many conserved quantities that in some cases enable us to construct multi-soliton solutions \cite{maddocks_stability_1993}. Here we will only discuss the more obvious  conserved quantities associated with the systems' global symmetries. The resulting constants of the motion  will not only be useful in the numerical calculation but also characterize the general physical properties of our model.

The Hamiltonian is time-translation invariant, hence the Hamiltonian itself as the time-translation generator is conserved. 
Also the invariance of Hamiltonian under global rotation ensures that each component $i=1,2,3$ of the total spin 
$$
M_i \stackrel{\rm def}{=} \int_{-\infty}^{\infty} m_i(x)\,dx
$$
is conserved.

A space-translation invariant continuous spin chain should possess a conserved momentum, but a rotationally invariant expression for the generator of space translations was long missing. This issue was elucidated by Haldane who showed \cite{haldane_geometrical_1986} that the quantum-mechanical generator of finite translations is the exponential of a Wess-Zumino term and is only well defined for discrete translations past an integer number of spin-$1/2$'s. If a local expression is required it necessarily involves a monopole gauge field in spin-space and the position of the Dirac string breaks rotational invariance. A similar issue arises in the purely classical chain. It is nonetheless useful in numerical calculations to introduce the quantity 
\begin{equation}
P = \int_{-\infty} ^{\infty} \vec{A}( \vec{m} ) \cdot \partial_x \vec{m} \,\, dx 
\end{equation}
where ${\bf A}$ is a monopole gauge field. Provided ${\bf m}(x)$ avoids the Dirac string at the south pole ${\bf m}=-\hat{\bf z}$, we can take
\begin{equation}
\vec{A} = \frac{(-\hat{\bf z}) \times \vec{m}}{1 - (-\hat{\bf z} \cdot \vec{m})}. 
\end{equation}
If we treat ${\bf m}(x)$ as a curve on the unit sphere parameterized by $x$, then $P$ is the area of the region enclosed by that curve on the side without the south pole. 

We then find that $\partial_x \vec{m} = \{\vec{m}, P\}$ for this restricted class of configurations. Because of the $4\pi$ ambiguity in defining the ``area enclosed'' by $\Gamma$ it is only the quantities 
 \begin{equation}
 T_a = e^{ i aP}, \quad a/2 \in {\mathbb Z},
 \end{equation}
 that are well-defined for general configurations.

\section{Single-speed Solutions}
\label{sec:single-speed}
We begin by seeking a single-soliton solution to \eqref{equ:EOM} that moves at a fixed velocity $v$, and so is of the form $\vec{m}(x,t) = \vec{m}(x-vt)$. The equation of motion then reduces to a non-linear ordinary differential equation
\begin{equation}
\label{equ:travel-wave}
v \frac{d \vec{m}}{dx} = {\bf m}\times \frac{d \vec{m}_{\mathcal{H}}}{dx}
\end{equation}
or, equivalently,
\begin{equation}
v \frac{d \vec{m}}{dx} = {\bf m}\times\left( \frac{d \vec{m}}{dx}\right)_{\mathcal{H}}.
\end{equation}

It is not obvious that equation \eqref{equ:travel-wave} possesses any interesting soliton-like solutions, so we first made a numerical search.

We followed the strategy used by Tjon and Wright \cite{tjon_solitons_1977} to find the Landau-Lifshitz soliton. 

We note that equation \eqref{equ:travel-wave} can be rewritten in term of the Poisson bracket as
\begin{equation}
 (\partial_t + v\partial_x)\vec{m} = \{\vec{m}, H+vP\}  = 0,
\end{equation}
and also note this same Poisson bracket equation gives the stationary points of the functional $H[\vec{m}] + vP[\vec{m}]$ subject to the variation on sphere. This because a variation a functional $I[\vec{m}]$ under $\delta \vec{m} =  \vec{m} \times \vec{\eta}$ is given by 
\begin{eqnarray}
\delta I = \int \left( \frac{\delta I}{\delta \vec{m}} \times \vec{m} \right) \cdot \vec{\eta}\,dx=\int \{{\bf m},I\}\cdot {\bm \eta} \,dx . 
\end{eqnarray}
Thus requiring $\delta I = 0$ for all ${\bm \eta}$ is the same as requiring $\{\vec{m}, I \}=0$. 
Consequently all the single-speed solutions are stationary points of $I_0[\vec{m}] = H[\vec{m}] + v P [\vec{m}]$.

Numerically accessible stationary points should be extrema\cite{dennis_numerical_1996}, so it is convenient to modify the functional $I_0$ and include a penalty term that makes the functional positive definite. If we take $I[\vec{m}] =  H + c_1 ( P - P_0)^2 $, at least one stationary configuration can be found by minimizing the functional and so solving
\begin{equation}
\label{equ:E-min}
 \{ \vec{m}, I\} = ( \partial_t + 2c_1 ( P - P_0) \partial_x ) \vec{m } = 0.
\end{equation}
As momentum is a conserved quantity, equation \eqref{equ:E-min} is identical to equation \eqref{equ:travel-wave} once the velocity is identified as 
\begin{equation}
\label{equ-v-E-min} 
v = 2c_1 ( P - P_0).
\end{equation}

A na{\"i}ve gradient descent to a minimum would take,
\begin{equation}
\vec{m}_{t+1} = \vec{m}_t + \vec{h}_{\text{eff}} \Delta t, \qquad   \vec{h}_{\text{eff}} = -\frac{\delta  I(\vec{m})}{\delta \vec{m}} 
\end{equation}
where the gradient $\vec{h}_{\text{eff}}$ is the effective field(mean field) produced by all the other spins and any external field. However, this strategy does not preserve the unit length of the spin field.

Instead, we project the gradient in the direction perpendicular to $\vec{m}$ in each updating step\cite{claas_abert_efficient_2014,berkov_solving_1993} and take 
\begin{equation}
\vec{m}_{t+\Delta t} = \vec{m}_t + [\vec{h}_{\text{eff}} - (\vec{h}_{\text{eff}} \cdot \vec{m}) \vec{m}] \Delta t.
\end{equation}
In the $\Delta t \rightarrow 0$ limit, this becomes the Gilbert damping equation \cite{gilbert_phenomenological_2004, hickey_origin_2009,lakshmanan_fascinating_2011},
\begin{equation}
\label{eq:Gilbert}
\frac{\partial \vec{m}}{\partial t} = -\vec{m} \times ( \vec{m} \times \vec{h}_{\text{eff}} ).
\end{equation}
Both the Gilbert damping equation \eqref{eq:Gilbert} and our generalized Landau-Lifshitz equation \eqref{equ:EOM} are time evolution equations of the type,
\begin{equation}
\frac{\partial \vec{m}}{\partial t } = \omega[ \vec{m} ] \times \vec{m }
\end{equation}
We wish to solve them accurately together while preserving the condition $|\vec{m}|^2 = 1$. 

Among many numerical schemes proposed for Landau-Lifshitz dynamics\cite{berkov_solving_1993, e_numerical_2000,banas_numerical_2005, cimrak_survey_2007} we found a mid-point finite difference to be the most suitable. This sets 
\begin{equation}
\frac{\vec{m}_{t+\Delta t} - \vec{m}_{t}}{\Delta t} = \vec{\omega}[\vec{m}_{t+ \frac{1}{2} \Delta t}] \times \frac{\vec{m}_{t+\Delta t} + \vec{m}_{t}}{2}
\end{equation}
and automatically preserves the length. In spite of being an implicit method, it can be made explicit by use of a predictor-corrector procedure.

We found that the relaxation process works cleanly and nearly all smooth initial conditions converged to a single-speed solution. The strength of the residual effective field $|\vec{h}_{\text{eff}}|$ is a measure of the final numerical error and is of order $10^{-6}$. We can also time-evolve the numerical final solution under equation \eqref{equ:EOM} and confirm that it moves at the calculated constant velocity with an error of order $10^{-5}$. 
\begin{widetext}
\begin{figure}[h]
 \centering
  \begin{tabular}{@{}p{0.45\linewidth}@{\quad}p{0.45\linewidth}@{}}
    \subfigimg[width=\linewidth]{\small{a)}}{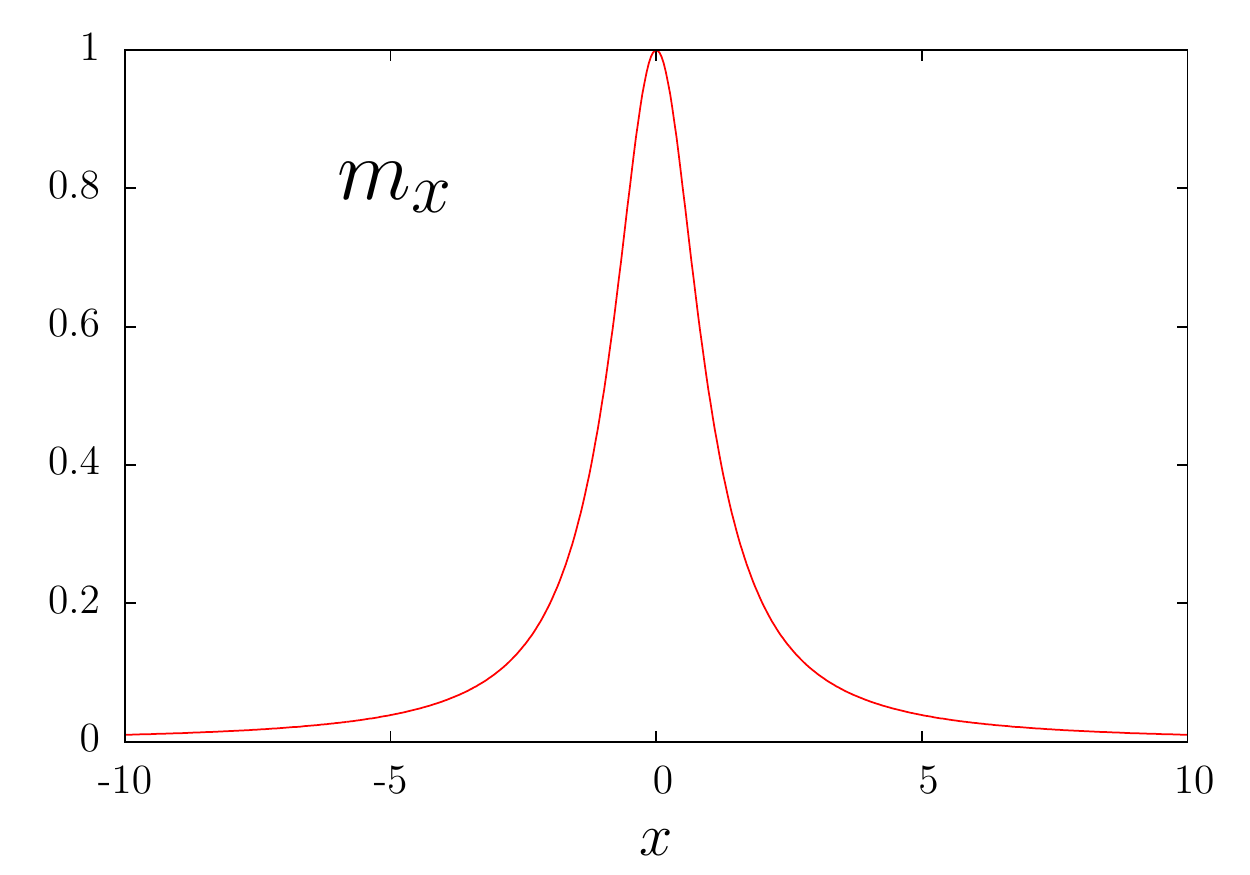} &
    \subfigimg[width=\linewidth]{\small{b)}}{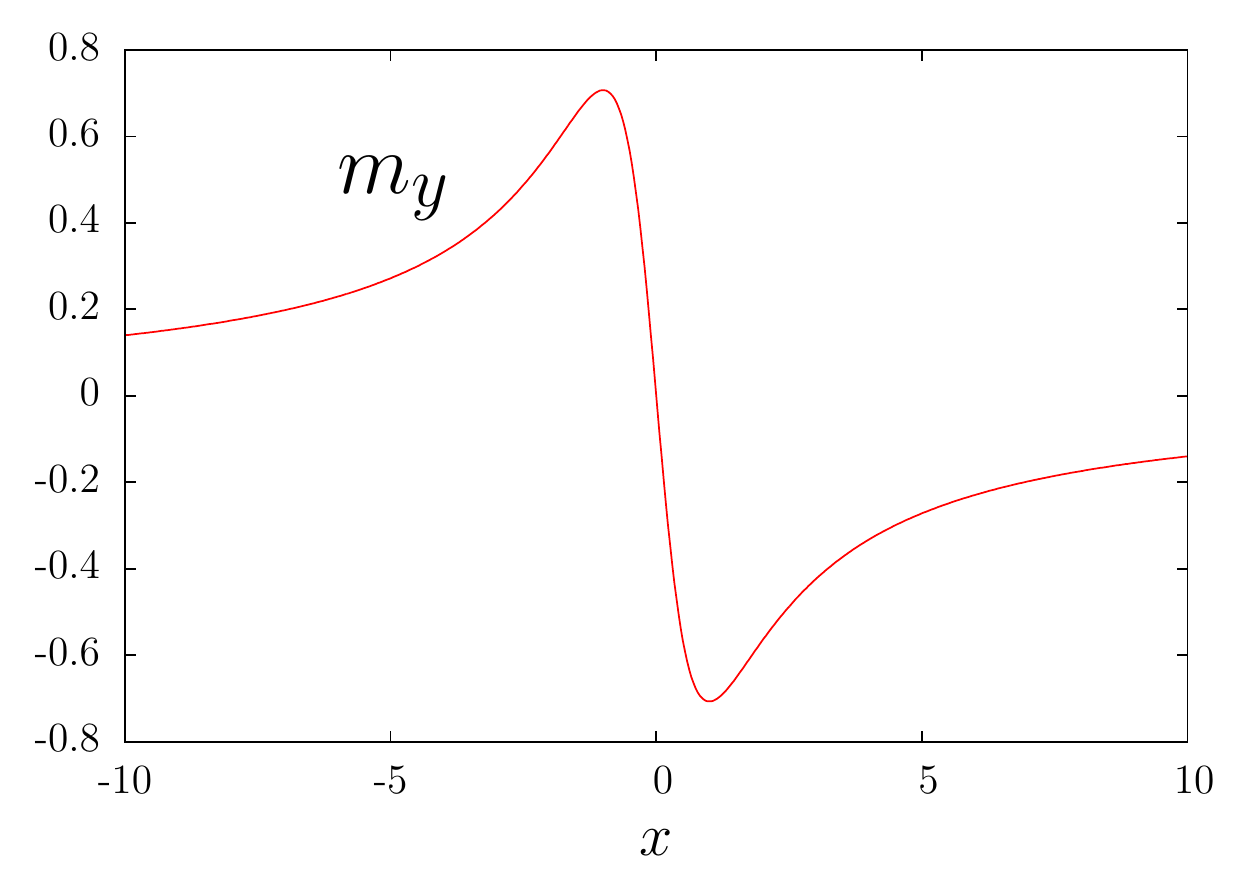} \\
    \subfigimg[width=\linewidth]{\small{c)}}{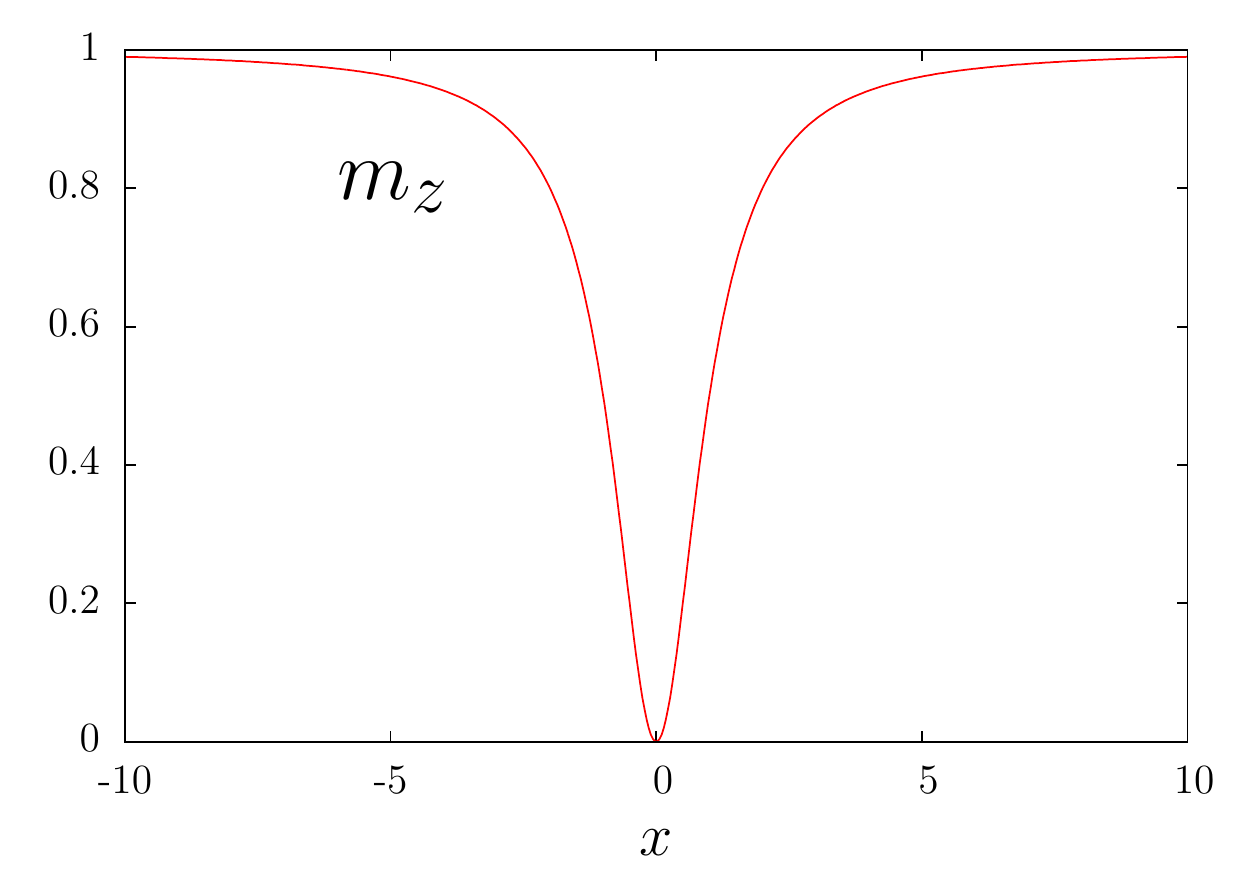} &
    \subfigimg[width=\linewidth]{\small{d)}}{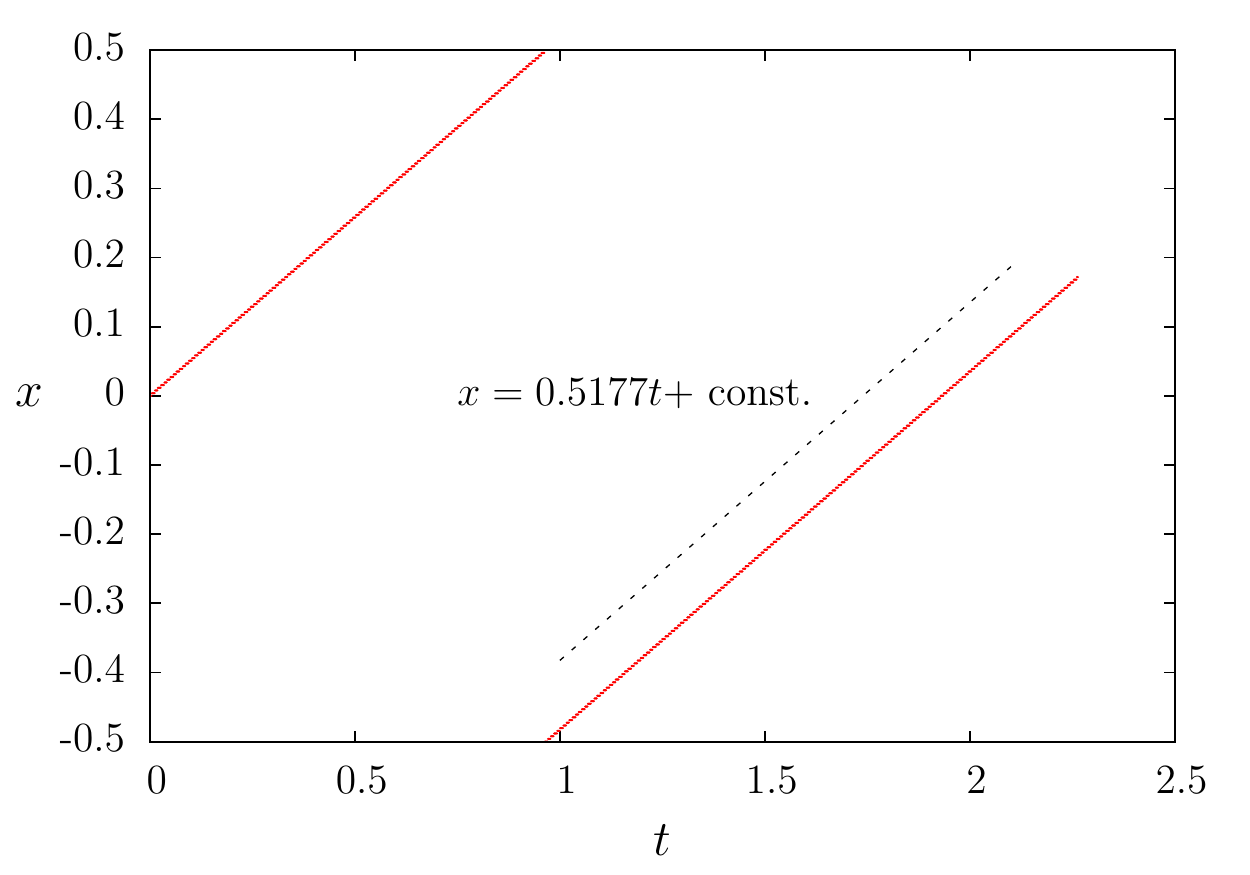}
  \end{tabular}
  \caption{A typical set of numerical results obtained by variational method. a),b) and c) are the three components of the single soliton solutions respectively. d) shows the trajectory of the minimal of $m_z$ in $x\in[-0.5, 0.5]$. A least square fit of line gives $v_{\text{measure}} = 0.5177$, while inserting the numerical value of $P$ and $P_0$ for this configuration into equation \eqref{equ-v-E-min} gives $v = 0.5176$.}
  \label{fig-single-pole-eg}
\end{figure}

\end{widetext}
Figure \ref{fig-single-pole-eg} demonstrates a typical set of data and fitting result. The measured velocity in the evolution agrees with the coefficient in equation \eqref{equ-v-E-min},  and so confirms the existence of the single soliton solution. 

Having reassured ourselves that there was at least one stable single soliton solution, we sought an analytic solution to \eqref{equ:travel-wave} by exploiting the fact that if a smooth real function $u(x)$ obeys some mild condition --- such as that it lies in some $L^p({\mathbb R}), p > 1$ for example --- and set $v(x)= u_{\mathcal H}$, then the function,
\begin{equation}
f(z) = \frac{1}{2\pi i} \int_{-\infty}^{\infty} \frac{u(x)+iv(x)}{x-z} dx
\end{equation}
is analytic in the upper half plane ${\mathbb H}$, tends to zero at infinity there, and has boundary value on the real axis   
\begin{equation}
f(x) =u(x)+iv(x).
\end{equation}
We therefore introduce a complex vector field $\vec{M}(z)$ whose components are analytic functions on the upper half plane and whose boundary values on the real axis are
\begin{equation}
\label{eq:def-M}
\vec{M}(x) = \vec{m}(x) + i\vec{m}_{\mathcal{H}}(x).
\end{equation}
From the traveling-wave equation \eqref{equ:travel-wave}, the real-axis normalization $|{\bf m}|^2=1$, and various Hilbert transform identities, we can deduce several results 
\begin{enumerate}
\item A decomposition of $\partial_x \vec{m}_{\mathcal{H}}$: 
\begin{equation}
\label{EQ:first-identity}
 \partial_x \vec{m}_{\mathcal{H}} = - v\vec{m} \times \partial_x \vec{m} +  \sqrt{1-v^2} |\partial_x \vec{m} | \vec{m}.  \qquad x\in \mathbb{R}
\end{equation}
\item Two orthogonality relations: 
\begin{equation}
\partial_x \vec{m}\cdot \partial_x \vec{m}_{\mathcal{H}} =0, \quad  \partial^2_x \vec{m}\cdot \partial^2_x \vec{m}_{\mathcal{H}} = 0.\qquad   x\in \mathbb{R}
\end{equation}
\item An analytic-function version of the orthogonality relations:
\begin{equation}
\label{eqn:cplx-orth}
 \partial_z \vec{M} \cdot \partial_z \vec{M} = 0,\quad \partial^2_z \vec{M} \cdot \partial^2_z \vec{M} = 0 \qquad  z \in  \mathbb{H}.
\end{equation}
\end{enumerate}
For proofs of these results see appendix \ref{app:equiv-forms}.

The first identity (\ref{EQ:first-identity}) tells us that a solution can only exist when the velocity is less than unity. Further, the complex orthogonal relation is very powerful. It simplifies the problem by trading the non-local Hilbert transform for the analytic function $\vec{M}$. Although initially derived by assuming that $z$ lies on the real axis, both left-hand sides of \eqref{eqn:cplx-orth} are analytic functions of $z$, and so both  equations must also  hold in the entire upper half plane $\mathbb{H}$. We can now solve the extended \eqref{eqn:cplx-orth} by expressing $M_3$ in terms of $M_1$ and $M_2$(Surprisingly, the solution implies $(\partial_z^n \vec{M})^2 =0$ for all $n > 2$),
\begin{eqnarray}
&&( \partial_z^2 M_1 )^2 + ( \partial_z^2 M_2 )^2 + (\partial_z \sqrt{- (\partial_z M_1)^2 - ( \partial_z M_2)^2})^2 = 0 \\
\implies && \partial_z (\ln \frac{\partial_z M_1}{\partial_z M_2}) =0 \\
\implies && \partial_z M_2 = c_2 \partial_z M_1 \quad \text{similarly} \quad  \partial_z M_3 = c_{3} \partial_z M_1 \\
\implies && 1 + c_2^2 + c_3^2 = 0 .
\end{eqnarray}

The result is that the three components $\partial_z \vec{M}$ are proportional to each other. 

After using the global rotational symmetry, the general solution $\vec{M}(z)$ can be parameterized by a single analytic function $g(z)$ and velocity $v = \cos \theta$, as
\begin{equation}
\label{eq:parameterization}
\vec{M}(z) = \left(-\frac{\sin 2\theta}{2}( g -1), i \sin \theta ( g - 1 ), 1 + \sin^2  \theta (g-1 )\right)
\end{equation}
The point-wise constraint $|\vec{m}(x)|^2 = 1$ that must hold on the real axis, together with  boundary conditions (values of $\vec{m}(\pm \infty)$) further dictate that
\begin{equation}
g(x) \bar{g}(x) = 1 \qquad g( x = \pm \infty ) = 1.
\end{equation}

A general solution can now be obtained by applying two-dimensional potential theory. The analyticity of the real and imaginary parts of the meromorphic functions $\Phi(z) = -\frac{1}{2\pi}\ln g(z) = u + iv$ implies that they satisfy the Poisson equation. They can therefore be regarded as the electrostatic potentials produced by charges located at zeros of $g(z)$. The condition $g(x) \bar{g}(x) = 1$ sets the boundary conditions $u(y= 0 ) = 0$. Hence $u(x,y)$ is just the potential of a semi-infinite metallic slab and obeys
\begin{equation}
-\nabla^2 u\big|_{\mathbb{H} / \{\text{zeros of } g\}}  = 0 \qquad u(y =0 ) = 0.
\end{equation}
By the method of images, we can remove the boundary condition at the price of placing an image charge at $z_i$ in the lower half plane for each charge located at $\bar{z}_i$. A general finite-charge solution is therefore
\begin{equation}
\label{EQ:pole-ansatz}
g(z) = \prod_{i=1}^N  \left[\frac{z - \bar{z}_i}{z-z_i} \right]. 
\end{equation}
For each such meromorphic $g(z)$, a spin field solution moving with velocity $v = \cos \theta $ is then
\begin{equation}
  \vec{m}_{\text{sol}} = (-\sin( 2\theta)(g_R-1)/2  , - \sin \theta g_I , 1 + \sin^2 \theta  (g_R-1)).
\end{equation}
It is remarkable that these single-speed solutions possess such a rich structure!

The reasoning leading to this general single-speed solution is quite subtle, so as a reality check we have confirmed --- both by substituting the proposed solution into the equation of motion and numerically --- that we have indeed found multi-lump single-speed solutions to \eqref{equ:travel-wave}. Because of the periodic boundary conditions used in our numerical work, only $g(z)$ with periodic poles are candidates for comparison. The simplest case is the periodic version of single pole solution, 
\begin{equation}
g(z) = \prod_{n = - \infty}^{\infty} \frac{z-n -ia }{z-n + ia } = \frac{\sin \pi( z- ia)}{\sin \pi( z+ ia ) }
\end{equation}
which reduces to $\frac{z-ia}{z+ia}$ when $z \ll a$. Setting $a$ and velocity $v$ as fitting parameters, the least square fit shows that the numerical and analytical results match perfectly with only a residue of order $10^{-5}$, which is roughly the order of numerical error. The fitting velocity for the configuration shown in figure \ref{fig-single-pole-eg} is $0.5176$, the difference is less than $10^{-6}$. 

\section{Numerical Multi-soliton Solutions}
\label{sec:multi-soliton}

The existence of single-speed multi-lump solutions suggests that there will be multi-soliton solutions that move at different speeds. We have not, however,  been able to generalize the pole ansatz (\ref{EQ:pole-ansatz}) so as to decouple the speeds and so find such solutions analytically. 

\begin{widetext}
\begin{figure}[h]
\begin{minipage}[t]{0.49\linewidth}
\centering
\includegraphics[width=\textwidth]{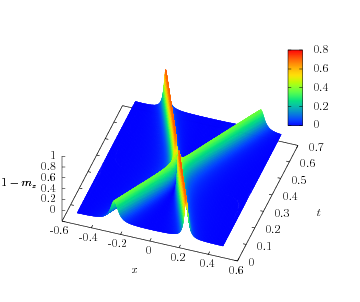}
\caption{\footnotesize Collisions of two single solitons. The displayed amplitude is $1 - m_z$ for better visualization. The velocities for the right-going and left-going solitons are $0.9$ and $-0.8$ respectively.}
\label{num_col.png}
\end{minipage}
\hfill
\begin{minipage}[t]{0.49\linewidth}
\centering
\includegraphics[width=\textwidth]{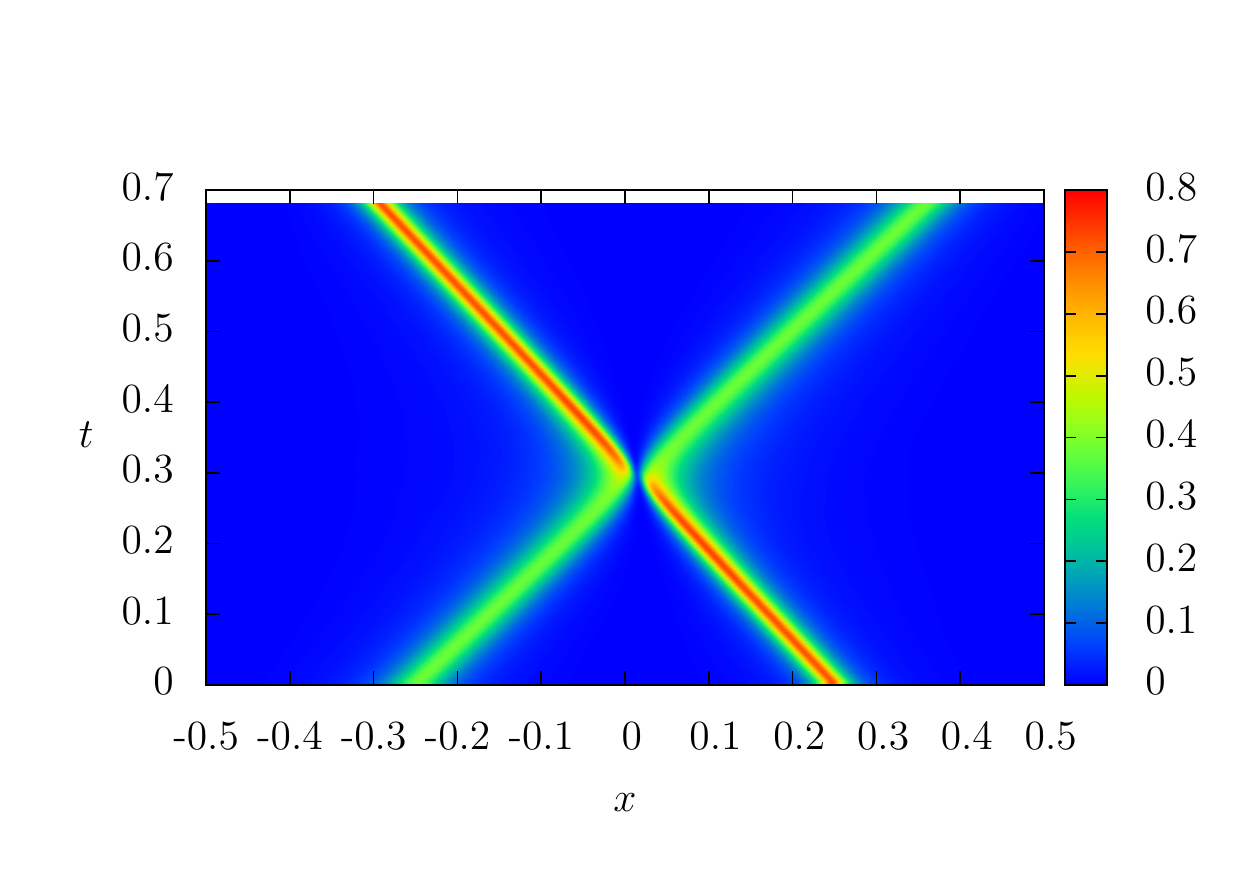}
\caption{\footnotesize Color map representation of the left figure. The larger amplitude soliton suffers less than the smaller one, so the time lag and advance are insignificant except for the region close to the center. The smaller starts to deviate earlier, but in the end either have net time lags.}
\label{num_col_proj.pdf}
\end{minipage}
\end{figure}
\end{widetext}

We therefore arranged several single-speed single-lump solutions some distance apart in the hope of approximating the initial conditions of a multi-soliton collision. We then evolved the spin field according to \eqref{equ:EOM}. Typical soliton interaction behaviors are shown in figure \ref{num_col.png} and \ref{num_col_proj.pdf}. The two solitons not only retain their initial profiles, but also appear to have zero time lag in the asymptotic region, as if the collision never occurred. 

\begin{widetext}
\begin{figure}[ht]
\begin{minipage}[t]{0.49\linewidth}
\centering
\includegraphics[width=\textwidth]{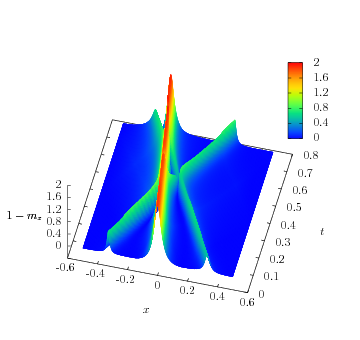}
\caption{\footnotesize Collisions of three single solitons. The velocities are $0.8$, $-0.25$ and $-0.85$ from left to right in the initial state. The three-soliton collision in this figure can be viewed as a "superposition" of several 2-body collisions. }
\label{num_3body.png}
\end{minipage}
\hfill
\begin{minipage}[t]{0.49\linewidth}
\centering
\includegraphics[width=\textwidth]{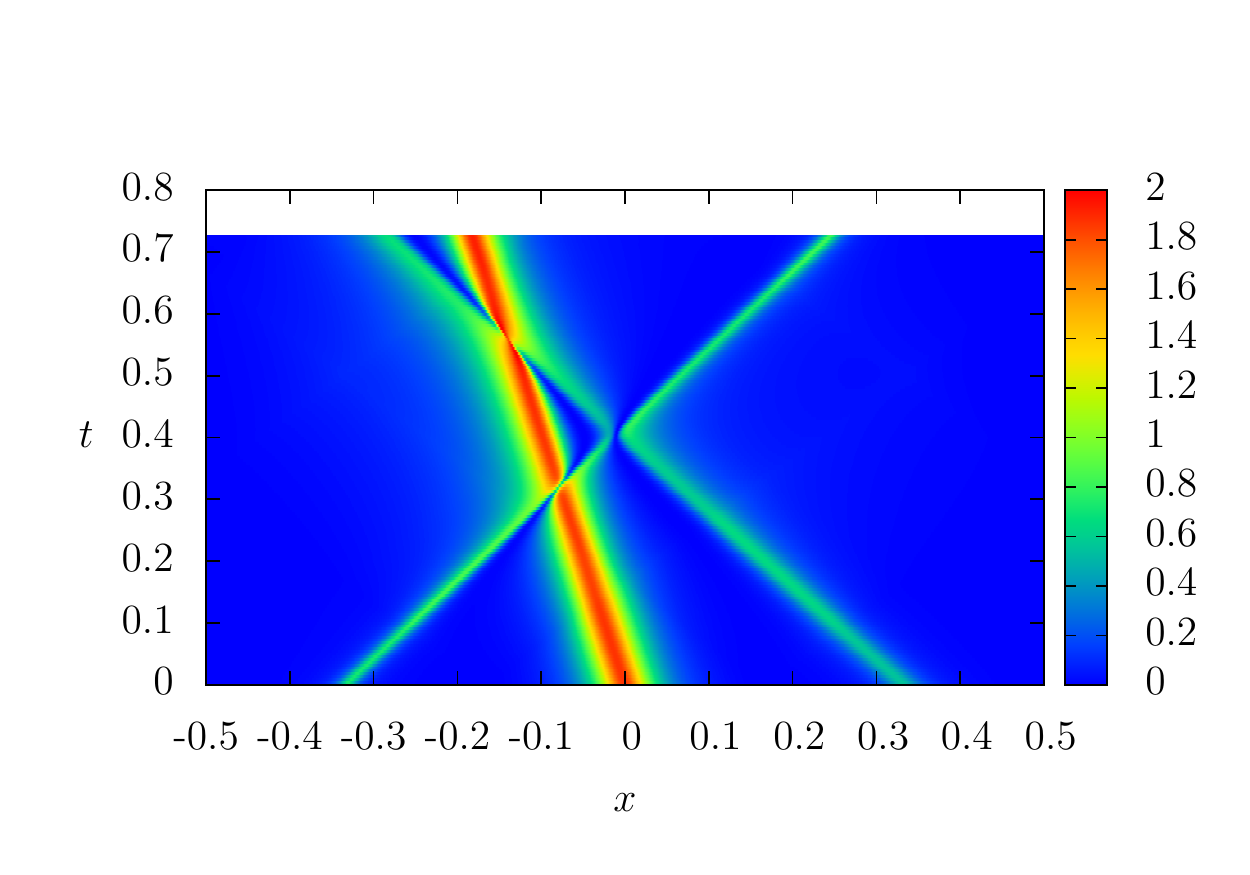}
\caption{\footnotesize Color map representation of the left figure. }
\label{num_3body_proj.pdf}
\end{minipage}
\end{figure}

\end{widetext}

Collisions of three solitons are shown in figure \ref{num_3body.png} and \ref{num_3body_proj.pdf}. Their behaviors are qualitatively the same as the superposition of three 2-body collisions. 

\section{Discussion}
\label{sec:discussion}

Through numerical and analytical calculations, we obtained a large family of single-speed multi-lump solutions to the generalized Landau-Lifshitz equation (\ref{equ:EOM}). The speed is always less than $1$ and serves as a parameter that controls the amplitude. The shape of each lump (defined as the full width at half maximum for each Lorentzian solution ) can be different depending upon the imaginary part of the pole.

One curious feature of the single-speed solutions is that the total energy and momentum are completely ignorant about the positions of the poles in $g(z)$. 

Consider the energy for a general spin field where $\vec{M} = ( f_1, f_2, 1+ f_3)$ with three possibly different analytic functions as its components. Each component $f_i$ maps the upper half-plane $\mathbb{H}$ to a region $D_i$. One can show that, 
\begin{equation}
E = \frac{1}{2i} \int \overline{\vec{M}} \cdot \partial_z \vec{M}  dz  =  \frac{1}{2i} \sum_{i=1}^3 \oint_{\partial D_i}  \bar{f}_i df_i = \sum_{i=1}^3 \int_{D_i} \frac{d\bar{f}_i df_i }{2i}.
\end{equation}
Hence the energy is the sum of the oriented areas swept out by the three analytic functions. 

For the single-speed solutions this gives 
\begin{equation}
\label{eq:energy}
E = 2\sin^2 \theta \int_{\text{unit disk}} \frac{d\bar{g} dg }{2i} = 2\pi N \sin^2 \theta  =  2\pi N (1-v^2),
\end{equation}
which is proportional to the winding number $N$ of $g$. 

Similarly the momentum $P$ is the solid angle enclosed by $\vec{m}(x)$\cite{haldane_geometrical_1986}. Hence it should depend only on the global property of the spin field. For the single-speed solution, it is also proportional to $N$,
\begin{equation}
\label{eq:momentum} 
  P  =  2\pi N ( 1 - \cos \theta ) =  2\pi N( v - 1 )
\end{equation}

The mysteries of these neat results can partly be elucidated by the geometry of the parameterization \eqref{eq:parameterization}. By taking a scalar product, we find the projection of $\vec{m}$ on the titled unit vector $\vec{n} = (\sin \theta, 0, \cos \theta)$ is the constant velocity  $\cos \theta$. So the trajectory of the tip of $\vec{m}$ on a unit sphere is in fact a small circle. Thus a more nature parameterization is to take $\vec{n}$ as the $z$ axis, and $\vec{m}$ to be
\begin{equation}
\label{eq:new-para}
\vec{m} = \Big(g_R\sin \theta,  g_I\sin \theta, \cos \theta  \Big)
\end{equation}
where $g = g_R + i g_I$ is the same analytic function defined in equation \eqref{EQ:pole-ansatz}. The two parameterizations of $\vec{m}$ are related through a rotation about the $y$ axis and thus are equivalent. Nonetheless the geometric meaning is clearer in equation \eqref{eq:new-para}. 
\begin{figure}[h]
\begin{minipage}[t]{0.47\linewidth}
\centering
\includegraphics[width=0.8 \textwidth]{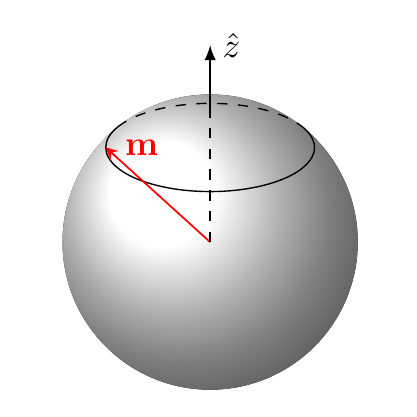}
\caption{ The single-speed solution $\vec{m}(x)$ viewed as a parametric curve of $x$ on unit sphere. When $x$ runs along the real axis, the tip of $\vec{m}$ traces out a circle on constant latitude $z = v = \cos \theta $ and repeats $N$ times for an $N$-lump solution. }
\label{sphere.pdf}
\end{minipage}
\hfill
\begin{minipage}[t]{0.47\linewidth}
\end{minipage}
\end{figure}
As shown in figure \ref{sphere.pdf}, for a $N$-lump single-speed solution, the tip of $\vec{m}$ traces out a small circle on the constant latitude plane $z = v = \cos \theta $ and repeats with possibly different paces $N$ times on this circle. The momentum $2N\pi( 1- \cos \theta )$ calculated in equation \eqref{eq:momentum} is actually the area of $N$ spherical caps. Furthermore, the image of the analytic function $M_x$ on the complex plane is a circle congruent to small circle in figure \ref{sphere.pdf}. Consequently, the area swiped by $M_x$ is area of $N$ of those small circles, which is $N \pi \sin^2 \theta$. The same is true for $M_y$. Taking into account that $M_z$ is constant, the total oriented area is thus $2\pi N \sin^2 \theta$ as we calculated before in the total energy equation \eqref{eq:energy}.

It now becomes apparent that all the dynamics of single-speed solution is constrained on a circle, which collapses the possibly existing multi-soliton space to the boring single-speed solution space. Generalizing the $U(1)$ group element $g$(on the real axis) in the single-speed solutions to the one in a larger group(e.g. $SU(2)$) is necessary to accommodate the numerically-found multi-solitons.

\section{Summary and Conclusions}
\label{sec:summary}

We have proposed a classical version of a continuum Haldane-Shastry spin model with the Hamiltonian $H = \int_{-\infty}^{\infty} dx \vec{m} \cdot \partial_x \vec{m}_{\mathcal{H} }$. The non-local Hilbert term is a consequence the of long-range interactions in the original quantum model, and the equation of motion contains a Hilbert transform that is similar to that in the integrable Benjamin-Ono equation. Motivated by the results of Polychronakos on the continuum Calogero-Sutherland model and of Abanov and Wiegmann\cite{abanov_quantum_2005} on the Benjamin-Ono equation, we conjectured that the model is integrable and support multi-soliton solutions. 

We numerically sought for and found single-soliton solutions that move with constant velocity $|v|<1$. We then found an analytic form for a large class of single-speed solutions by rewriting the equations in terms of analytic functions on the upper half-plane $\mathbb{H}$. These solutions contain multiple lumps with possibly different widths. We found that the energy and momentum of a $N$-lump solution are proportional to the topological invariant winding number $N$. From the geometric point of view, it is the consequence of the circular trajectory shown in figure \ref{sphere.pdf}.

We also performed numerical soliton-soliton collision experiments and found that the single-lump solitons survive multiple collisions unscathed. There appears to be no asymptotic time lag after the collisions, and this, and the role of the winding number $N$, suggests that there should be some clever transformation to a system of non-interacting particles. At the moment we have no idea of how to find this transformation.     

\section{Acknowledgements} 
This work is supported by the National Science Foundation under grant number NSF-DMR-13-06011. T.Z. would like to thank Lei Xing for his code for evolving the spinful Calogero-Sutherland model. 

\appendix

\section{Two methods to calculate hydrodynamic limit}

\label{app:hydro-limit}

The double summation we are going to do is
\begin{equation}
  H = \rho^2 \sum_{i\ne j} \frac{1 - \vec{m}_i \cdot \vec{m}_j }{(i-j)^2} = \rho^2 \sum_i \vec{m}_i \cdot \vec{S}_i
\end{equation}
where $\vec{S}_{i} = \sideset{}{'}\sum_{j = -\infty}^{\infty} \frac{ \vec{m}_i  -   \vec{m}_j}{ (i-j)^2 }$ is the inner sum. The total sum and the inner sum $\vec{S}_i$ are both absolutely convergent, so it is legitimate to reorder and do the inner sum first. 

Our first approach is to use a special Euler-Maclaurin summation technique introduced in reference \cite{stone_classical_2008}. Euler-Maclaurin formula converts a discrete sum into an integral of the interpolation function $f(x)$
\begin{equation}
\sum_{n = a+1} ^{b} f(n) = \int_a^bf(x)  dx + \sum_{k=0}^{p} \frac{B_k}{k!}f^{(k-1)}(x)\Big|^b_a + R_p, \quad a,b \in \mathbb{Z}, a<b.
\end{equation}
If the interpolation function is smooth and has exponential type less than $2\pi$, then the only corrections are derivative at boundaryies and the remainder term
\begin{equation}
R_p = (-1)^p \int_{a}^{b} \frac{1}{p!} B_p( x - \left\lfloor x \right\rfloor) f^{(p)}(x)dx,
\end{equation}
where Bernoulli numbers $B_k$ and Bernoulli polynomials $B_p(x)$ are accompany with the $p$th derivative. 

Due to the singularity at $i = j$, one can not na\"ively apply it to the inner sum $\vec{S}_{i}$ . However it is possible to introduce a counter term as in \cite{stone_classical_2008} to remove the singularity without affecting the result. The regulated summand is $\vec{f}_{ij} =\frac{\vec{m}_i -  \vec{m}_j}{ (i-j)^2 } - \frac{\vec{m}_i'}{i-j}$ so that
\begin{eqnarray}
\vec{S}_{i} =&& \lim_{N\rightarrow \infty} \sideset{}{'}\sum_{j = -N}^{j = N }\vec{f}_{ij} =  \lim_{N \rightarrow \infty}  \int_{-N}^{N}d\nu \vec{f}_{i\nu}  \\
&&- \lim_{\nu \rightarrow i }\vec{f}_{i\nu}
+ \lim_{N \rightarrow \infty} \sum_{k=1}^{p} \frac{B_k}{k!} \vec{f}^{(p)}_{i\nu}\Big|^{N}_{-N}  + R_p.
\end{eqnarray}
We only focus on the soliton solutions, which vanishes asymptotically $f_{ij}^{(p)}(\pm \infty) = 0$. Hence the derivative corrections at $\pm N$ go to zero for fixed order $p$. The remainder term has a bound
\begin{equation}
R_p \le \frac{2\zeta(2p)}{(2\pi)^{2p}} \int_{-\infty}^{\infty} | f^{(p)}( x) | dx,
\end{equation}
for the finite domain summation, the remainder term can be arbitrarily small by taking a large enough order $p$. For the infinite domain, in most cases the remainder term can also be neglected for functions whose exponential type is less than $2\pi$( It is however not necessarily zero, see an example of applying this formula to the 1d monoatomic gas partition function in chapter 7 of the book\cite{graham_concrete_1994}).

Granted that we can neglect the remainder term, the discrete sum has only two terms left
\begin{equation}
\label{eqn:eu-ma-si}
\vec{S}_i = \pi {\mathcal{H}} ( \vec{m}'_i ) + \frac{1}{2} \vec{m}_i''
\end{equation}
where the Hilbert transform comes from the integral, the double derivative comes from ${\bf f}_{ii}$. Completing the sum over $i$ gives the total energy,
\begin{eqnarray}
H  &=& \rho^2 \sum_{i=-\infty}^{\infty} \vec{S}_i \cdot \vec{m}_i = \rho^2 \int_{-\infty}^{\infty} d\mu \bigg\{\frac{1}{2} \vec{m}(\mu)  \vec{m}''(\mu) \\
&&+ \vec{m}(\mu) \cdot P \int_{-\infty}^{\infty}d \nu\big[ \frac{\vec{m}(\mu) -  \vec{m}(\nu)}{ (\mu-\nu)^2 }\big]\bigg\}\\
\label{eq:total-energy}
 &=& \rho^2 \int_{-\infty}^{\infty} d\mu [ - \frac{1}{2} \vec{m}'^2 + \pi \vec{m} \cdot \vec{m}'_{\mathcal{H}}].
\end{eqnarray}

An alternative way to do the sum is the Fourier interpolation. The advantage of this particular interpolation is that the discrete sum is automatically equal to the integral, because the zeroth component of Fourier series is defined to be the sum! 

The interpolation starts with expanding a periodic function $f(n)$ by its discrete Fourier series
\begin{equation}
f(n) = \sum_{\frac{k}{2\pi}=-\frac{N}{2}+1}^{\frac{N}{2}} f_k e^{ i k n },
\end{equation}
and then extending the definition of Fourier series for integer number $n$ to real number $x$
\begin{equation}
f(x) = \sum_{\frac{k}{2\pi}=-\frac{N}{2}+1}^{\frac{N}{2}} f_k e^{ i k x }.
\end{equation}
What immediately follows is that
\begin{equation}
\sum_{n = 1}^{N} f(n) = \int_0^N f(x) dx.
\end{equation}

Now take any component of $\vec{m}$ to be a periodic function $f(n)$, the corresponding component in $\vec{S}_l$ is then
\begin{eqnarray}
\sideset{}{'} \sum_{j = -\infty}^{\infty} \frac{g(l) - g(j) }{ (j-l) ^2} &=& \sum_{\frac{k}{2\pi}=-\frac{N}{2}+1}^{\frac{N}{2}}  f_k e^{ikl} \sideset{}{'}\sum_{j = -\infty}^{\infty}  \frac{1 - e^{i k (j-l)} }{ (j-l)^2} \\
&=& \sum_{\frac{k}{2\pi}=-\frac{N}{2}+1}^{\frac{N}{2}}  f_k e^{ikl} \sideset{}{'}\sum_{n = -\infty}^{\infty}  \frac{1 - e^{i k n} }{ n^2} 
\label{equ:math-SE}
\end{eqnarray}
Identity (A.4) in the reference \cite{loganayagam_anomaly/transport_2012} provides a closed form expression for the summation over $n$. In the notations of this appendix, the identity is
\begin{equation}
\text{Li}_n(e^{i k} ) + (-1)^n \text{Li}_n(e^{-i k } ) = - \frac{(2\pi i )^ n}{n!} B_n ( \frac{k}{2\pi})  
\end{equation}
where $\text{Li}_n(x)$ is the polylogarithm function
\begin{equation}
\text{Li}_s(z) = \sum_{n=1}^{\infty} \frac{z^n}{n^s}.
\end{equation}
which is equivalent to 
\begin{equation}
\label{id:bernoulli}
\sideset{}{'}\sum_{n = -\infty}^{\infty}  \frac{e^{i k n} }{ n^s}  = - \frac{s!}{(2\pi i)^s} B_s( \frac{k}{2\pi} ).
\end{equation}
We need the case of $s = 2$, so that for $0<|k| < 2\pi$
\begin{eqnarray}
 \sideset{}{'} \sum_{-\infty}^{\infty} \frac{e^{i k n }-1 }{ n^2} &=& \frac{1}{2} (  k^2 - 2\pi |k|  ).
\end{eqnarray}
Therefore
\begin{eqnarray}
\sideset{}{'} \sum_{j = -\infty}^{\infty} \frac{g(l) - g(j) }{ (j-l) ^2}&=&  \sum_{\frac{k}{2\pi}=-\frac{N}{2}+1}^{\frac{N}{2}}  f_k e^{ikl}  \frac{1}{2} (  -k^2 + 2\pi |k|  ), 
\end{eqnarray}
and the inner sum becomes
\begin{equation}
\sideset{}{'} \sum_{j = -\infty}^{\infty} \frac{g(l) - g(j) }{ (j-l) ^2} = \frac{1}{2} f'' + \pi  \mathcal{H}(f')\big|_{x = l}.
\end{equation}

The result of $S_l$ agrees with \eqref{eqn:eu-ma-si}. Moreover, there is no approximation involved in replacing the outer sum by integral for Fourier interpolated functions. Therefore the total energy expression $H$ in equation \eqref{eq:total-energy} becomes exact in this method.

\section{Equivalent forms of traveling wave equation}
\label{app:equiv-forms}

This appendix presents the algebras to deduce several equivalent forms of the traveling wave equation \eqref{equ:travel-wave}.

We begin by showing equation \eqref{equ:travel-wave} is equivalent to a decomposition
\begin{equation}
\label{eqn:decomp}
\partial_x \vec{m}_{\mathcal{H}} = - v\vec{m} \times \partial_x \vec{m} +  \sqrt{1-v^2} \big|\partial_x \vec{m} \big| \vec{m} .
\end{equation}

Assuming $\partial_x \vec{m}$ is nowhere zero, the set $\{ \vec{m}, \partial_x \vec{m}, \vec{m} \times \partial_x\vec{m}\}$ forms an orthogonal basis. So $\partial_x \vec{m}_{\mathcal{H}}$ can be expanded as a linear superposition of them. Taking the scalar product of the traveling wave equation \eqref{equ:travel-wave} with $\partial_x \vec{m}_{\mathcal{H}}$, we find $\partial_x \vec{m} \cdot \partial_x \vec{m}_{\mathcal{H}} = 0$. Hence there is no $\partial_x \vec{m}$ component in the decomposition of $\partial_x \vec{m}_{\mathcal{H}}$
\begin{equation}
\label{eq:alpha-beta}
\partial_x \vec{m}_{\mathcal{H}} = \alpha \vec{m} \times \partial_x \vec{m} +  \beta \vec{m} .
\end{equation}
Taking a cross product of \eqref{eq:alpha-beta} with $\vec{m}$, the consistency with equation \eqref{equ:travel-wave} requires $\alpha = -v$.
To determine $\beta$, we expand $\mathcal{H}(\partial_x \vec{m} \cdot \partial_x \vec{m})$ by the Hilbert transform identity for the product $\mathcal{H}(fg) = f_\mathcal{H} g + f g_\mathcal{H}  + \mathcal{H}(f_\mathcal{H} g_\mathcal{H})$\cite{king_hilbert_2009} to get
\begin{equation}
\label{eq:hilbert-id}
\mathcal{H}( \partial_x \vec{m} \cdot \partial_x \vec{m} - \partial_x \vec{m}_{\mathcal{H}} \cdot \partial_x \vec{m}_{\mathcal{H}}) = 0.
\end{equation}
Whatever in the braces is a constant, which is set to zero by boundary conditions. So the two vectors have equal lengths
\begin{equation}
\label{eq:equal-length}
\partial_x \vec{m} \cdot \partial_x \vec{m} = \partial_x \vec{m}_{\mathcal{H}} \cdot \partial_x \vec{m}_{\mathcal{H}},
\end{equation}
The ``equal-length'' condition together by squaring equation \eqref{eq:alpha-beta} allows us to compute $\beta$ 
\begin{eqnarray}
|\partial_x \vec{m}_{\mathcal{H}} | = v^2 |\partial_x \vec{m}|^2 + \beta^2 
 \implies \beta = \pm \sqrt{1 - v^2} |\partial_x \vec{m}| .
\end{eqnarray}
The positive definiteness of the total energy $\int_{-\infty}^{\infty} \vec{m} \cdot \partial_x \vec{m}_{\mathcal{H}}  = \int_{-\infty}^{\infty}\beta d x >0$ picks out the positive $\beta$ branch. We arrive at the decomposition \eqref{eqn:decomp}. Since the converse is trivial, this completes the proof of their equivalence. A byproduct is the that the speed is always less than $1$.

Furthermore, the decomposition we just derived is also equivalent to the two orthogonality relations
\begin{equation}
\partial_x \vec{m}\cdot \partial_x \vec{m}_{\mathcal{H}} = 0 \qquad  \partial^2_x \vec{m}\cdot \partial^2_x \vec{m}_{\mathcal{H}} = 0.
\end{equation}
We have proved the first in equation \eqref{eq:equal-length}. The second can be verified by differentiating the decomposition relation \eqref{eqn:decomp} and doing a scalar product
\begin{eqnarray}
\partial_x^2 \vec{m} \cdot \partial_x^2 \vec{m}_{\mathcal{H}} &=& \big[- v\vec{m} \times \partial^2_x \vec{m} +  \sqrt{1-v^2} \partial_x (|\partial_x \vec{m} | \vec{m})  \big] \cdot \partial_x^2 \vec{m} \\
&=& \sqrt{1-v^2}  \frac{1}{|\partial_x \vec{m} |}  (\partial_x \vec{m} \cdot \partial_x^2 \vec{m}) \\
&&\big[ ( \vec{m} \cdot \partial_x^2 \vec{m}) + |\partial_x \vec{m}|^2 \big] \\
&=& 0.
\end{eqnarray}
So the second derivative terms are also perpendicular to each other. Similar to the derivation in equation \eqref{eq:hilbert-id}, we can even prove an analogous ``equal-length'' relation
\begin{equation}
\label{eq:equal-length2}
\partial_x^2 \vec{m} \cdot \partial^2_x \vec{m} = \partial_x^2 \vec{m}_{\mathcal{H}} \cdot \partial_x^2 \vec{m}_{\mathcal{H}}.
\end{equation}

On the other hand, given these two orthogonality relations, we can also deduce equation \eqref{equ:travel-wave}. The first orthogonality relation tells us that $\partial_x \vec{m}_{\mathcal{H}} $ can only have the following decomposition(note that $\alpha$ may be position dependent)
\begin{equation}
 \partial_x \vec{m}_{\mathcal{H}} = \alpha \vec{m} \times \partial_x \vec{m} \pm  \sqrt{1 - \alpha^2 } |\partial_x \vec{m} | \vec{m} .
\end{equation}
From our calculation above, $\partial^2_x \vec{m}_{\mathcal{H}} \cdot \partial_x^2 \vec{m} =0$ if $\alpha$ is constant. So the scalar product must be proportional to the derivative of $\alpha$ 
\begin{eqnarray}
\partial_x^2 \vec{m} \cdot \partial_x^2 \vec{m}_{\mathcal{H}} &=& \partial_x \alpha ( \vec{m} \times \partial_x \vec{m} \mp  \frac{\alpha}{\sqrt{1-\alpha^2 }}  |\partial_x \vec{m} | \vec{m} ) \cdot \partial_x^2 \vec{m}\\
&=& \frac{\partial_x \alpha}{\alpha} \partial_x^2 \vec{m} \cdot \partial_x^2 \vec{m}_{\mathcal{H}} \mp \frac{\partial_x \alpha}{\alpha \sqrt{1-\alpha^2}} |\partial_x \vec{m} |\vec{m}\cdot \partial_x^2 \vec{m}\\
&=& \pm \frac{\partial_x \alpha}{\alpha \sqrt{1-\alpha^2}} |\partial_x \vec{m} |^3 = 0  ,
\end{eqnarray}
the second orthogonality relation restricts $\alpha$ to be a constant. 
 
In summary, these derivations rewrite the traveling wave equation in completely different forms without losing any information. Moreover the two orthogonality relations and subsequent ``equal-length'' relations \eqref{eq:equal-length} and \eqref{eq:equal-length} are the real and imaginary parts of the following complex orthogonality relations
\begin{equation}
\label{equ:complex-ortho}
 \partial_z \vec{M} \cdot \partial_z \vec{M} = 0 \quad  \partial^2_z \vec{M} \cdot \partial^2_z \vec{M} = 0 \qquad \forall z \in  \mathbb{R}
\end{equation}
where $\vec{M}$ is defined in equation \eqref{eq:def-M}. They are zero on the real axis, and therefore also zero in the entire upper half plane. This is  the key to completely solve the single-speed sector. 
  
\bibliographystyle{unsrt}
\bibliography{HS_model}

\begin{thebibliography}{10}

\bibitem{filippov_great_2010}
Alexandre~T. Filippov.
\newblock The {Great} {Solitary} {Wave} of {John} {Scott} {Russell}.
\newblock In {\em The {Versatile} {Soliton}}, Modern {Birkhäuser} {Classics},
  pages 23--37. Birkhäuser Boston, 2010.

\bibitem{stegeman_optical_1999}
George~I. Stegeman and Mordechai Segev.
\newblock Optical {Spatial} {Solitons} and {Their} {Interactions}:
  {Universality} and {Diversity}.
\newblock {\em Science}, 286(5444):1518--1523, 1999.

\bibitem{bjorkholm_cw_1974}
J.~E. Bjorkholm and A.~A. Ashkin.
\newblock cw {Self}-{Focusing} and {Self}-{Trapping} of {Light} in {Sodium}
  {Vapor}.
\newblock {\em Phys. Rev. Lett.}, 32(4):129--132, 1974.

\bibitem{strecker_formation_2002}
Kevin~E. Strecker, Guthrie~B. Partridge, Andrew~G. Truscott, and Randall~G.
  Hulet.
\newblock Formation and propagation of matter-wave soliton trains.
\newblock {\em Nature}, 417(6885):150--153, 2002.

\bibitem{cooper_propagating_1999}
N.~R. Cooper.
\newblock Propagating {Magnetic} {Vortex} {Rings} in {Ferromagnets}.
\newblock {\em Phys. Rev. Lett.}, 82(7):1554--1557, 1999.

\bibitem{sutcliffe_vortex_2007}
Paul Sutcliffe.
\newblock Vortex rings in ferromagnets: {Numerical} simulations of the
  time-dependent three-dimensional {Landau}-{Lifshitz} equation.
\newblock {\em Phys. Rev. B}, 76(18):184439, 2007.

\bibitem{niemi_leapfrogging_2014}
Antti~J. Niemi and Paul Sutcliffe.
\newblock Leapfrogging vortex rings in the {Landau}–{Lifshitz} equation.
\newblock {\em Nonlinearity}, 27(9):2095, 2014.

\bibitem{sutherland_beautiful_2004}
Bill Sutherland.
\newblock {\em Beautiful {Models}: 70 {Years} of {Exactly} {Solved} {Quantum}
  {Many}-{Body} {Problems}}.
\newblock World Scientific, 2004.

\bibitem{polychronakos_physics_2006}
Alexios~P. Polychronakos.
\newblock Physics and {Mathematics} of {Calogero} particles.
\newblock {\em Journal of Physics A: Mathematical and General},
  39(41):12793--12845, 2006.

\bibitem{polychronakos_waves_1995}
Alexios~P. Polychronakos.
\newblock Waves and {Solitons} in the {Continuum} {Limit} of the
  {Calogero}-{Sutherland} {Model}.
\newblock {\em Physical Review Letters}, 74(26):5153--5157, 1995.

\bibitem{andric_solitons_1995}
I.~Andrić, V.~Bardek, and L.~Jonke.
\newblock Solitons in the {Calogero}-{Sutherland} {Collective}-{Field} {Model}.
\newblock {\em Physics Letters B}, 357(3):374--378, 1995.

\bibitem{haldane_exact_1988}
F.~D.~M. Haldane.
\newblock Exact {Jastrow}-{Gutzwiller} resonating-valence-bond ground state of
  the spin-(1/2 antiferromagnetic {Heisenberg} chain with 1/${\mathrm{r}}^{2}$
  exchange.
\newblock {\em Phys. Rev. Lett.}, 60(7):635--638, 1988.

\bibitem{shastry_exact_1988}
B.~Sriram Shastry.
\newblock Exact solution of an \textit{{S}} =1/2 {Heisenberg} antiferromagnetic
  chain with long-ranged interactions.
\newblock {\em Phys. Rev. Lett.}, 60(7):639--642, 1988.

\bibitem{haldane_spinon_1991}
F.~D.~M. Haldane.
\newblock "{Spinon} gas" description of the \textit{{S}} =1/2 {Heisenberg}
  chain with inverse-square exchange: {Exact} spectrum and thermodynamics.
\newblock {\em Phys. Rev. Lett.}, 66(11):1529--1532, 1991.

\bibitem{polychronakos_lattice_1993}
Alexios~P. Polychronakos.
\newblock Lattice integrable systems of {Haldane}-{Shastry} type.
\newblock {\em Phys. Rev. Lett.}, 70(15):2329--2331, 1993.

\bibitem{talstra_integrals_1995}
J.~C. Talstra and F.~D.~M. Haldane.
\newblock Integrals of motion of the {Haldane}-{Shastry} model.
\newblock {\em J. Phys. A: Math. Gen.}, 28(8):2369, 1995.

\bibitem{bernard_yang-baxter_1993}
D.~Bernard, M.~Gaudin, F.~D.~M. Haldane, and V.~Pasquier.
\newblock Yang-{Baxter} equation in long-range interacting systems.
\newblock {\em J. Phys. A: Math. Gen.}, 26(20):5219, 1993.

\bibitem{polychronakos_exchange_1992}
Alexios~P. Polychronakos.
\newblock Exchange operator formalism for integrable systems of particles.
\newblock {\em Phys. Rev. Lett.}, 69(5):703--705, 1992.

\bibitem{benjamin_internal_1967}
T.~Brooke Benjamin.
\newblock Internal waves of permanent form in fluids of great depth.
\newblock {\em Journal of Fluid Mechanics}, 29(03):559--592, 1967.

\bibitem{ono_algebraic_1975}
Hiroaki Ono.
\newblock Algebraic {Solitary} {Waves} in {Stratified} {Fluids}.
\newblock {\em J. Phys. Soc. Jpn.}, 39(4):1082--1091, 1975.

\bibitem{joseph_multi-soliton-like_1977}
R.~I. Joseph.
\newblock Multi-soliton-like solutions to the {Benjamin}–{Ono} equation.
\newblock {\em Journal of Mathematical Physics}, 18(12):2251, 1977.

\bibitem{tjon_solitons_1977}
J.~Tjon and Jon Wright.
\newblock Solitons in the continuous {Heisenberg} spin chain.
\newblock {\em Phys. Rev. B}, 15(7):3470--3476, 1977.

\bibitem{fogedby_solitons_1980}
H.~C. Fogedby.
\newblock Solitons and magnons in the classical {Heisenberg} chain.
\newblock {\em J. Phys. A: Math. Gen.}, 13(4):1467, 1980.

\bibitem{lakshmanan_fascinating_2011}
M.~Lakshmanan.
\newblock The {Fascinating} {World} of {Landau}-{Lifshitz}-{Gilbert}
  {Equation}: {An} {Overview}.
\newblock {\em Philosophical Transactions of the Royal Society A: Mathematical,
  Physical and Engineering Sciences}, 369(1939):1280--1300, 2011.

\bibitem{betchov_curvature_1965}
Robert Betchov.
\newblock On the curvature and torsion of an isolated vortex filament.
\newblock {\em Journal of Fluid Mechanics}, 22(03):471--479, 1965.

\bibitem{hasimoto_soliton_1972}
Hidenori Hasimoto.
\newblock A soliton on a vortex filament.
\newblock {\em Journal of Fluid Mechanics}, 51(03):477--485, 1972.

\bibitem{lamb_solitons_1976}
G.~L. Lamb.
\newblock Solitons and the {Motion} of {Helical} {Curves}.
\newblock {\em Phys. Rev. Lett.}, 37(5):235--237, 1976.

\bibitem{lakshmanan_continuum_1977}
M.~Lakshmanan.
\newblock Continuum spin system as an exactly solvable dynamical system.
\newblock {\em Physics Letters A}, 61(1):53--54, 1977.

\bibitem{zakharov_equivalence_1979}
V.~E. Zakharov and L.~A. Takhtadzhyan.
\newblock Equivalence of the nonlinear {Schrödinger} equation and the equation
  of a {Heisenberg} ferromagnet.
\newblock {\em Theor Math Phys}, 38(1):17--23, 1979.

\bibitem{takhtajan_integration_1977}
L.~A. Takhtajan.
\newblock Integration of the continuous {Heisenberg} spin chain through the
  inverse scattering method.
\newblock {\em Physics Letters A}, 64(2):235--237, 1977.

\bibitem{lakshmanan_dynamics_1976}
M.~Lakshmanan, Th.~W. Ruijgrok, and C.~J. Thompson.
\newblock On the dynamics of a continuum spin system.
\newblock {\em Physica A: Statistical Mechanics and its Applications},
  84(3):577--590, 1976.

\bibitem{bikbaev_landau-lifshitz_2014}
R.~F. Bikbaev, A.~I. Bobenko, and A.~R. Its.
\newblock Landau-{Lifshitz} equation, uniaxial anisotropy case: {Theory} of
  exact solutions.
\newblock {\em Theor Math Phys}, 178(2):143--193, 2014.

\bibitem{maddocks_stability_1993}
John~H. Maddocks and Robert~L. Sachs.
\newblock On the stability of {KdV} multi-solitons.
\newblock {\em Comm. Pure Appl. Math.}, 46(6):867--901, 1993.

\bibitem{haldane_geometrical_1986}
F.~D.~M. Haldane.
\newblock Geometrical {Interpretation} of {Momentum} and {Crystal} {Momentum}
  of {Classical} and {Quantum} {Ferromagnetic} {Heisenberg} {Chains}.
\newblock {\em Phys. Rev. Lett.}, 57(12):1488--1491, 1986.

\bibitem{dennis_numerical_1996}
J.~Dennis and R.~Schnabel.
\newblock {\em Numerical {Methods} for {Unconstrained} {Optimization} and
  {Nonlinear} {Equations}}.
\newblock Classics in {Applied} {Mathematics}. Society for Industrial and
  Applied Mathematics, 1996.

\bibitem{claas_abert_efficient_2014}
Gregor~Wautischer Claas~Abert.
\newblock Efficient {Energyminimization} in {Finite}-{Difference}
  {Micromagnetics}: {Speeding} up {Hysteresis} {Computations}.
\newblock {\em Journal of Applied Physics}, 116(12), 2014.

\bibitem{berkov_solving_1993}
D.~V. Berkov, K.~Ramstöcck, and A.~Hubert.
\newblock Solving {Micromagnetic} {Problems}. {Towards} an {Optimal}
  {Numerical} {Method}.
\newblock {\em phys. stat. sol. (a)}, 137(1):207--225, 1993.

\bibitem{gilbert_phenomenological_2004}
T.L. Gilbert.
\newblock A phenomenological theory of damping in ferromagnetic materials.
\newblock {\em IEEE Transactions on Magnetics}, 40(6):3443--3449, 2004.

\bibitem{hickey_origin_2009}
M.~Hickey and J.~Moodera.
\newblock Origin of {Intrinsic} {Gilbert} {Damping}.
\newblock {\em Physical Review Letters}, 102(13), 2009.

\bibitem{e_numerical_2000}
W.~E and X.~Wang.
\newblock Numerical {Methods} for the {Landau}--{Lifshitz} {Equation}.
\newblock {\em SIAM J. Numer. Anal.}, 38(5):1647--1665, 2000.

\bibitem{banas_numerical_2005}
L’ubomír Baňas.
\newblock Numerical {Methods} for the {Landau}-{Lifshitz}-{Gilbert} {Equation}.
\newblock In {\em Numerical {Analysis} and {Its} {Applications}}, number 3401
  in Lecture {Notes} in {Computer} {Science}, pages 158--165. Springer Berlin
  Heidelberg, 2005.

\bibitem{cimrak_survey_2007}
Ivan Cimrák.
\newblock A {Survey} on the numerics and computations for the
  {Landau}-{Lifshitz} equation of micromagnetism.
\newblock {\em ARCO}, 15(3):1--37, 2007.

\bibitem{abanov_quantum_2005}
Alexander~G. Abanov and Paul~B. Wiegmann.
\newblock Quantum {Hydrodynamics}, the {Quantum} {Benjamin}-{Ono} {Equation},
  and the {Calogero} {Model}.
\newblock {\em Phys. Rev. Lett.}, 95(7):076402, 2005.

\bibitem{stone_classical_2008}
Michael Stone, Inaki Anduaga, and Lei Xing.
\newblock The classical hydrodynamics of the {Calogero}-{Sutherland} model.
\newblock {\em Journal of Physics A: Mathematical and Theoretical},
  41(27):275401, 2008.

\bibitem{graham_concrete_1994}
Ronald~L. Graham, Donald~Ervin Knuth, and Oren Patashnik.
\newblock {\em Concrete {Mathematics}: {A} {Foundation} for {Computer}
  {Science}}.
\newblock Addison-Wesley, 1994.

\bibitem{loganayagam_anomaly/transport_2012}
R.~Loganayagam and Piotr Surówka.
\newblock Anomaly/transport in an {Ideal} {Weyl} gas.
\newblock {\em J. High Energ. Phys.}, 2012(4):1--34, 2012.

\bibitem{king_hilbert_2009}
Frederick~W. King.
\newblock {\em Hilbert {Transforms}}, volume~1.
\newblock Cambridge University Press, Cambridge, 2009.

\end{thebibliography}

\end{document}